\documentstyle[12pt]{article}
\input{epsf.tex}
\def\lsim{\mathrel{\rlap {\raise.5ex\hbox{$<$}}
{\lower.5ex\hbox{$sim$}}}}
\def\gsim{\mathrel{rlap {\raise.5ex\hbox{$>$}}
{\lower.5ex\hbox{$sim$}}}}

\begin{document}
\begin{titlepage}
\begin{flushright}
CERN-TH/96-54\ \\
hep-ph/9602389{\hskip.5cm}\\
\end{flushright}
\begin{centering}
\vspace{.3in}
{\bf R-PARITY VIOLATION AND PECCEI-QUINN SYMMETRY IN GUTS} \\
\vspace{3 cm}
{K.TAMVAKIS$^{\ast}$}
\vskip 1cm
{\it Theory Division, CERN}\\
{\it 1211 Geneva 23, Switzerland}\\
\vspace{1.5cm}
{\bf Abstract}\\
\end{centering}
\vspace{.1in}
We address the question whether it is possible in GUTs to obtain 
R-parity violation with a large
\begin{math}{\Delta L{/}\Delta B}\end{math} hierarchy of strengths so that 
the proton is stable while phenomenologically interesting $L$-violation
is present. We consider versions of SU(5) with a built-in Peccei-Quinn
symmetry spontaneously broken at an intermediate scale.
The P-Q symmetry and the field content guarantee a large  suppression of
the effective $B$-violating terms by a factor
\begin{math}\Lambda^3_{PQ}/M_PM^2_X \end{math}
while the effective $L$-violating terms stay large.
\vspace{2cm}
\begin{flushleft} CERN-TH/96-  \\
February 1996\\
\end{flushleft}
\hrule width 6.7cm \vskip.1mm{\small \small \small $^\ast$\ On leave from
Physics Department,University of Ioannina,Ioannina GR45110,Greece.}
\end{titlepage}
\newpage

\begin{bf}1.Introduction.\end{bf}
 A straightforward supersymmetrization of the Standard Model\cite{NHK}
allows the existence of low dimension operators (D=4,5) that 
violate $B$- and $L$-number\cite{WSY}. The D=4 operators are usually 
avoided by imposing a discrete symmetry called R-parity\cite{ZHS} 
and defined as $R=(-1)^{3B+L+2S}$ with $S$ being the spin. Similarly,
the dangerous D=5 operators are eliminated by 
imposing a suitable symmetry. If this symmetry is broken 
at some intermediate scale $\Lambda$, these operators will be 
supressed by $\Lambda /M $,
$M$
being a large mass scale. The 
Peccei-Quinn \cite{PQ} symmetry 
proposed for the explanation of the vanishing vacuum angle 
theta is such a symmetry suitable for the suppression of the
$ B$-violating D=5 operators. Examples of GUTs incorporating 
a P-Q symmetry have been constructed\cite{GRH,HMY,HMT}.
 
If R-parity is not a symmetry of the Standard Model, then 
the superpotential should include(directly or effectively)the 
terms
\begin{equation}{\lambda _{ijk}l_il_je^c_k +
\lambda '_{ijk}d^c_il_jq_k +\lambda ''_{ijk}d^c_id^c_ju^c_k +
\epsilon_il_iH}\end{equation}
The indices are generation indices. The combination of the second 
and third term results in proton decay through squark exchange at an
unacceptable rate unless $\left|\lambda' \lambda''\right|\leq10^{-24}$.
If one is restricted 
within the Standard Model it is possible, addopting a phenomenological
attitude, to assume the existence of some of these couplings while 
forbidding the presense of others\cite{DRG}.
For example, setting $\lambda''_{ijk}=0$
 while keeping the rest leads to a number of $L$-violating phenomena.
 This is something that cannot be done in GUTs, at least in such a
straightforward fashion. For instance, in SU(5) all terms in (1) 
can arise from
\begin{equation}\lambda_{ijk}\phi_i(\overline{5} )\phi_j(\overline{5} )
\psi_k(10) +\epsilon_i \phi_i(\overline{5} ) H(5) \end{equation}
In SU(5) all couplings in (1) are related by
$\lambda''_{ijk} =\frac{1}{2}\lambda'_{ijk}=\lambda_{ijk}$
and should be present simultaneously.
Then, if R-parity is not an exact symmetry, a large
 hierarchy in $B$-versus $L$-violating strengths must be accounted for\cite{SV}.

Nevertheless, it is possible that these terms could be absent at 
the renormalizable level due to another symmetry, not directly 
related to R-parity, and show up as non-renormalizable effective
 interactions leading to small effective couplings suppressed by
 ratios of the breaking scale of this symmetry to some large mass 
scale. Note however that the required smallness of these couplings 
comes about almost exclusivelly from the need to suppress the
 $B$-violating interactions threatening the proton stability.
 $L$-violating couplings, if they were independent as in the
 Standard Model, they would not be so severely constrained. A model
 of effective R-parity violation would be phenomenologically
 interesting if it were characterized by an effective large 
$B$-versus $L$-violation disparity.

\begin{bf}2.Peccei-Quinn symmetry in SU(5) and R-parity violation.\end{bf} 
 A P-Q -symmetric version of the minimal supersymmetric SU(5)
 model can be constructed in a straightforward fashion at the
 expense of introducing an extra pair of Higgs pentaplets and 
singlets\cite{HMY}. The superpotential of the model is
$$W=h_{ij}\psi_i \psi_j H +f_{ij}\psi_i\phi_j \overline{H}
+\overline{H}'(M'+\lambda'\Sigma )H +\overline{H}(M''+\lambda'' \Sigma )H'
+f\overline{H}HP $$  
 \begin{equation}+f'\overline{H}'H'\overline{P} +(M/2)Tr(\Sigma^2)
+(\lambda /3)Tr(\Sigma^3) \end{equation}
The extra fields are the pentaplets $H'$,
$\overline{H}'$and the SU(5)-singlets $P$, $\overline{P}$.
The charges under $U(1)_{PQ}$are $\psi(1)$, $\phi(1)$,$H(-2)$,
$\overline{H}(-2)$,$H'(2)$, $\overline{H}'(2)$, $\Sigma(0)$, $P(4)$,
$\overline{P}(-4)$. In order to generate the 
required PQ-breaking we need to add to (3) suitable additional 
interactions among the singlets.
Couplings $hP \overline{P} X $ to another (neutral)
 singlet $X$ with a mass of $O(M_P)$,when $X$ is integrated out, lead to 
effective non-renormalizable terms
$h^2(P\overline{P})^2/M $. Such a term would be 
sufficient to induce spontaneous breaking of the P-Q
symmetry\cite{MSY}, in conjunction with the standard soft
supersymmetry breaking terms
 in the potential
$m_0^2 (\left|P\right|^2 + \left|\overline{P}\right|^2)$
and $ m_0Ah^2(P\overline{P})^2/M + $ h.c   . 
The scale of P-Q breaking is $ \langle P\rangle=\langle\overline{P}
\rangle\equiv \mu=-m_0M(A/6h^2)(1+\sqrt{1-12/A^2})\simeq 10^{10}$--$10^{12}$
Gev. This range 
of values is compatible with astrophysical and cosmological
 bounds\cite{K}.

 R-parity, although not explicitely imposed, is an exact symmetry
 of the model even after P-Q spontaneous breaking. Although we 
cannot exclude that R-parity is indeed an exact symmetry it is
 more interesting to explore the possibility that additional 
interactions exist which ultimately lead to effective R-violating
 couplings among standard fields such as $\phi_i H$, $\phi_i H'$ and
$\psi_i \phi_j \phi_k$ .
 For instance, singlets carrying odd P-Q charge could couple to 
the above operators. A viable model however should predict also 
the neccessary suppression of these effective couplings. It is
 possible to costruct models in which the P-Q charges of the 
fields guarantee that the R-violating operators will appear at the
non-renormalizable level. As one of the possible classes of models 
that could be constructed, we shall consider a pair of singlets $S$,
$\overline{S}$
carrying P-Q charge 1/2. This choice of charge ensures the 
absence of renormalizable couplings to the other fields. Then, the 
R-violating term \begin{equation}\lambda_i \phi_i H S^2/M \end{equation}
is possible. A P-Q-breaking v.e.v.
 for $S$, $\overline{S}$ breaks R-parity and generates an effective Higgs-matter
 mixing through this term.
A v.e.v. $\overline{\mu}=\langle S\rangle=\langle\overline{S}\rangle\sim
\frac{m_0M}{\overline{h}^2} $, of the same order 
of the $P$ and $Q$ v.e.v., can be generated through the presence of 
a term $\overline{h}(S\overline{S})^2/M$ in conjunction with soft
supersymmetry breaking. All these terms can arise as
 effective interactions from couplings $\phi HY +SS\overline{Y}
+\overline{S}Y\overline{S} +S\overline{S}Z $ to singlets
$Z$,$Y$,$\overline{Y}$
having masses of the order of the Planck-mass, after they are
 integrated out.Higher order R-violating terms
\begin{equation} (\phi H')S^2\overline{P}/M^2\end{equation}
and
\begin{equation}(\psi\phi\phi)\overline{P}S^2/M^3 \end{equation} are also
 present but their suppression with  extra powers of the Planck
 mass makes them not relevant.Terms with $\Sigma $ insertions can
also be written down but they are suppressed by powers of $ M_X/M $.

\begin{bf}3.Higgs-matter mixing.\end{bf} 
 Taking into account the interactions in (3) and (4), the 
Higgs-pentaplet mass-matrices are
\begin{equation} M^{(2)}=\left[\begin{array}{rrr}
(f\mu)&M_2&0\\
\overline{M_2}&(f'\mu)&0\\
0&0&0
\end{array}\right]\end{equation} and 
\begin{equation}M^{(3)}=\left[\begin{array}{rrr}
(f\mu)&M_3&\epsilon\\
\overline{M_3}&(f'\mu)&0\\
0&0&0
\end{array}\right]\end{equation} 
in a $H_2$, $H'_2$ / $\overline{H_2}$, $\overline{H'_2}$,
$l_0$ and $H_3$, $H'_3$ / $\overline{H_3}$, $\overline{H'_3}$,
$d^c_0$  basis. The matter fields $l_0 $ and $d^c_0$
are the combinations appearing in the coupling (4)
\begin{equation} (\lambda_i \langle S\rangle^2/M)\phi_iH=\epsilon_i\phi_iH=
\epsilon(l_0H_2+d^c_0H_3)\end{equation} 
We have set $\epsilon_i=\lambda_i\langle S\rangle^2/M $ and
$\epsilon=(\sum\epsilon_i^2)^{1/2}$. Notice that $\epsilon $ is
of the order of $\lambda\mu^2/M$ , $\mu$
being the P-Q breaking scale set by the $\langle S\rangle$, $\langle P\rangle$
v.e.v's.

 The isodoublet mass-eigenvalues can be read-off from
$M^{(2)}(M^{(2)})^{\dagger}$. 
At this point we should impose the, inevitable, fine-tunning 
that will guarantee a light mass-eigenvalue. It is convenient 
to put it in the form of the condition
\begin{equation}(M_2\overline{M_2}-ff'\mu^2)^2=\epsilon^2(M_2^2+(f\mu)^2)
\end{equation} 
implying the appearence of a mass-eigenvalue of the order of $\epsilon$. The 
resulting eigenvalues are
\begin{equation}(m_2)_{+}=(M_2^2+\overline{M}_2^2+(f\mu)^2+(f'\mu)^2)^{1/2}
\end{equation} and \begin{equation}(m_2)_-=\epsilon\end{equation}
Note that $(m_2)_+$ is of the order of $\mu$ since the 
condition (10) amounts to requiring that $M_2$, $\overline{M}_2$
 are of that order.
 The combination \begin{equation}l=[(f'\mu)\overline{H}_2-(\overline{M}_2)
\overline{H}'_2 +(M_2^2+(f\mu)^2)^{1/2}l_0]/(m_2)_+\end{equation}
is massless. The intermediate mass
 isodoublets
 $\overline{H}_+$, $H_+$ will have an appreciable influence on the running
of gauge 
couplings. This is however within the limits allowed by existing
 data in correlation with proton decay\cite{HMY}. The colour-triplet 
eigenvalues are both of order $M_3$, $\overline{M}_3$. The combination
\begin{equation}d^c=N[d^c_0+\epsilon(M_3\overline{M}_3-ff'\mu^2)^{-1}((f'\mu)
\overline{H}_3-\overline{M}_3\overline{H}'_3)]\end{equation} is
 massless.

 The standard down-quark Yukawa interactions written in terms of
"mass-eigenstates"are 
\begin{equation} Y^{(d)}_i[(l_{0i}e^c_i+q'_id^c_{0i})\overline{H}_2
+(l_{0i}q'_i+u^c_id^c_{0i})\overline{H}_3]\end{equation} 
with $q'_i=(u_i,V_{ij}d_j)$ in terms of the Kobayashi-Maskawa matrix  $V_{ij}$. 
The combinations that mix with Higgses are $\epsilon_il_{0i}$ and
$\epsilon_id^c_{0i}$.In general, all $\epsilon_i$'s are non-zero. We could
always go to a new basis in which
 the combination $\epsilon_i\phi_i$ will define one family. For example,
$l_1=l_{01}$, $l_2=l_{02}$ and $l_3=\epsilon_il_{0i}/\epsilon$.The new
Yukawa's, according to (15) will
be $Y'_i=Y_i-Y_3\epsilon_i/\epsilon_3$ for i=1,2
 and $Y'_3=Y_3\epsilon/\epsilon_3$. Nevertheless, it might be
plausible\cite{SV}, and certainly 
simplifying, to assume a family hierarchy in $\epsilon_i$ proportional 
to the hierarchical structure of the $Y_i$'s. 
In that case we could consider in $\epsilon_i\phi_i$ only the contribution
 of the (dominant) third family.Therefore, we procceed by 
assuming that only the third family has an appreciable
 R-violating coupling.

 Substituting the expressions of $l_{03}$ , $d^c_{03}$ , $\overline{H}_2$
and $\overline{H}_3$  in terms of the light 
eigenstates, we obtain the leading order Yukawa coupling of 
the third generation $$Y^{(d)}_3[\frac{M_2}{\sqrt{M_2^2+(f'\mu)^2}}
(l_3\tau^c\overline{H}_-) + \frac{M_2^2+2(f\mu)^2}{\sqrt{(M_2^2+(f'\mu)^2)
(M_2^2+(f\mu)^2)}}(q'_3b^c\overline{H}_-)$$
 \begin{equation}-\frac{(f'\mu)M_2}{\sqrt{(M_2^2+(f'\mu)^2)
(M_2^2+(f\mu)^2)}}(q'_3b^cl_3) +\ldots]\end{equation} 
No $B$-violating coupling appears due to colour antisymmetry.
 In contrast, the $L$-violating coupling $q'_3b^cl_3$ appears with an 
O(1) coefficient. The Yukawa's of the other two generations are
\begin{equation}\sum_{i=1,2}Y^{(d)}_i[\frac{M_2^2+2(f'\mu)^2}
{\sqrt{(M_2^2+(f'\mu)^2)(M_2^2+(f\mu)^2)}}(l_ie^c_i+q'_id^c_i)
\overline{H}_- \end{equation}  $$  -\frac{(f'\mu)M_2}
{\sqrt{(M_2^2+(
f'\mu)^2)(M_2^2+(f\mu)^2)}}(l_ie^c_i+q'_id^c_i)l_3 +
(\epsilon(f'\mu)/M_3\overline{M}_3)(q'_il_i+u^c_id^c_i)b^c]$$
Note the presence of the $L$-violating interactions $\mu^cl_{\mu}l_{\tau}$ ,
$e^cl_el_{\tau}$ , $s^cq'_2l_{\tau}$ , $d^cq'_1b^c$    with 
O(1) couplings while the $B$-violating  operators $c^cs^cb^c$ , $u^cd^cb^c$ 
carry a
 drastic suppression factor $\epsilon(f'\mu)/M_3\overline{M}_3$ .
This is a rather small number
 of the orderd of $10^{-20}$ . This should be compared to the "direct"
 $B$-violating term $(\phi_i\phi_j\psi_k)\overline{P}S^2/M^3$
 which carries an even smaller coefficient
 of the order of $(\mu/M)^3$ .

 The above hierarchy of $L$-versus $B$-non-conservation is sufficient
 to guarantee a stable proton since \begin{equation}\lambda'\lambda''
\sim(m_{\mu}/v_1)^2\epsilon(f'\mu)/M_3\overline{M}_3\leq10^{-24}\end{equation}
 Nevertheless a 
number of procceces not respecting Lepton -number result 
from (17).The interaction $\nu_{\tau}b'b^c$ generates at one loop a mass 
for the $\tau$--neutrino, roughly $\frac{Y_b^2}{16\pi^2}
(m_b/\tilde{m}_b)^2Am_{3/2}$ which,
 being of the order of Mev,  is easily in agreement
 with existing cosmological bounds\cite{K,DPV}.

\begin{bf}4.Other models.\end{bf}
 In the P-Q SU(5) model that has been analyzed,the scale of
 P-Q breaking has been "naturally" determined by the other
 scales present ( $m_{3/2}$ , $M_P$ ) and by the particular form of the 
superpotential couplings of the fields dictated by the 
symmetries. The suppression of R-violating terms, as in the 
analogous suppression of D=5 operators that break Peccei-Quinn, is entirely
independent of the fine-tunning required for the
 triplet-doublet splitting. This is much more clear in the 
so-called missing-doublet SU(5) model\cite{MNT} endowed with a P-Q
 symmetry\cite{HMT}. This model has been constructed in order to
 avoid the fine numerical adjustment in the triplet-doublet
 mass splitting required in the minimal model.
 The superpotential is \begin{equation}W=\psi \psi H +\psi\phi\overline{H}+
\overline{\lambda}\overline{H}\Sigma\overline{\Theta} +\lambda H \Sigma\Theta
+\frac{M}{2}Tr(\Sigma^2)+\frac{h}{3}Tr(\Sigma^3)\end{equation}
$$ +\overline{\lambda}'\overline{H}'\Sigma \overline{\Theta}' +\lambda'H'
\Sigma\Theta'+M_1\Theta\overline{\Theta}'+M_2\Theta'\overline{\Theta}$$
The SU(5) and $U(1)_{PQ}$ quantum numbers of the fields are
$\psi(10,\alpha/2)$, $\phi(\overline{5},\beta/2)$, $H(5,-\alpha)$,
$\overline{H}'(\overline{5},\alpha)$, $\overline{H}
(\overline{5},-(\alpha+\beta)/2)$, $H'(5,(\alpha+\beta)/2)$,
$\Theta(50,\alpha )$, $\Theta'(50,-(\alpha+\beta))$, $\overline{\Theta}
(\overline{50},(\alpha+\beta)/2)$, $\overline{\Theta}'
(\overline{50},-\alpha)$, $\Sigma(75,0)$ .The masses $M_1$, $M_2$ are taken
to be of the order of
 the Planck-mass in order to avoid an increase of the gauge 
coupling beyond the perturbativity limit due to the presence 
of too many light fields. Integrating out the superheavy 50's
 we obtain the effective superpotential \begin{equation}\psi \psi H +
\psi\phi\overline{H}+H_3'\overline{H}_3M_3+H_3\overline{H}_3'\overline{M}_3
\end{equation} in which 
only the colour-triplets appear with masses $M_3=\lambda\overline{\lambda}'
\langle\Sigma\rangle^2/M_1$ , $\overline{M}_3=\lambda'\overline{\lambda}
\langle\Sigma\rangle^2/M_2$. 
Both these masses are slightly bellow the unification scale,
namely $10^{14}$--$10^{15}$Gev. There is no mass terms for the doublets as a 
consequence of the absence of direct mass terms for the pentaplets.

 In addition to the interactions appearing in (19), new interaction
 terms are possible if gauge-singlet fields,charged under 
P-Q are introduced. Being a little different than the case of 
the minimal P-Q SU(5), we introduce $P(-(3\alpha+\beta)/2)$ ,
$Q(3(3\alpha+\beta)/2)$ and $S((\alpha-\beta/2)/3)$. No other
renormalizable terms are possible with these fields except
\begin{equation}fP\overline{H}'H'\end{equation} Again, various
non-renormalizable interactions are present. 
They are \begin{equation}P^3Q/M+(\overline{H}H)P^2Q/M^2+\tilde{\lambda}_iS^3
(\phi_iH)/M^2\end{equation}
All these terms can be written down for charges defined for 
independent $\alpha$'s and $\beta$'s. This reflects the existence of two 
U(1)'s of which one can be broken by an extra interaction of
 leading non-renormalizable order 1/M among the fields $P$ , $Q$ , $S$ 
 that forces a relation among the phases. For example, the 
interaction $P^2QS/M$ enforces the, peculiar, phase relation $2\beta=-11\alpha$ .
 In any case,the breaking of the $U(1)_{PQ}$ procceeds in a similar
 way as in the minimal model coming out again in the range
$10^{10}$--$10^{12}$ Gev.

 The Higgs pentaplet mass-matrices are
\begin{equation}M^{(2)}=\left[\begin{array}{rrr}
\hat{\epsilon}&0&\epsilon\\0&(f\mu)&0\\0&0&0
\end{array}\right]\end{equation} and
\begin{equation}M^{(3)}=\left[\begin{array}{rrr}
\hat{\epsilon}&\overline{M}_3&\epsilon\\M_3&(f\mu)&0\\0&0&0
\end{array}\right]\end{equation}
in a $H_2$ , $H'_2$ / $\overline{H_2}$ , $\overline{H}'_2$ ,
$l_{03}$ and $H_3$ , $H'_3$ / $\overline{H}_3$ , $\overline{H}'_3$ ,
$d^c_{03}$ basis. Again for simplicity
 we have assumed that the R-non-conserving coupling is 
exclusively to the third generation.   The appearing
 parameters are $\epsilon=\tilde{\lambda}\langle S\rangle^3/M^2
\sim10^2$--$10^3$Gev, for an intermediate P-Q scale choice of $10^{11}$ Gev,
 and $\hat{\epsilon}=\langle P\rangle^2\langle Q\rangle/M^2$ ,
roughly of the same order. The doublet
 mass-matrix leads to eigenvalues $m_+^2=(f\mu)^2$ and $m_-^2=\epsilon^2+
\hat{\epsilon}^2$ . The combination $l_{\tau}=(\epsilon\overline{H}_2-
\hat{\epsilon}l_{03})/m_-$ is massless. The triplet eigenvalues are both of
order $M_3\sim\overline{M}_3$. The combination
 \begin{equation}b^c=N[\epsilon(f\mu)\overline{H}_3 -
\epsilon M_3\overline{H}'_3+(M_3\overline{M}_3-\hat{\epsilon}f\mu)d^c_{03}]
\end{equation}is massless. Expressing the 
down-quark Yukawa's in terms of eigenstates we obtain
\begin{equation}Y^{(d)}_3[l_{\tau}\tau^c\overline{H}_- +\frac{\hat{\epsilon}}
{\sqrt{\epsilon^2+\hat{\epsilon}^2}}(q'_3b^c\overline{H}_-)+
\frac{\epsilon}{\sqrt{\epsilon^2+\hat{\epsilon}^2}}(q'_3b^cl_{\tau})]
\end{equation} 
$$\sum_{i=1,2}Y^{(d)}_i[(q'_il_i+d^c_iu^c_i)b^c(\frac{\epsilon f\mu}{M_3^2})
(\frac{M_3\overline{M}_3}{M_3^2-\overline{M}_3^2}-(\frac{M_3}
{\overline{M}_3})^3)+(q'_id^c_i+l_ie^c_i)(\overline{H}_-\hat{\epsilon}+
l_{\tau}\epsilon)/\sqrt{\epsilon^2+\hat{\epsilon}^2}]$$
Again, the $\Delta B/\Delta L$  hierarchy is of order  $\epsilon\mu/M_3^2$
 and $\lambda'\lambda''\sim(m_{\mu}/v_1)^2(\epsilon f\mu/M_3^2)\leq10^{-24}$ .

\begin{bf}5.Brief summary.\end{bf}
 The $L$-violating couplings of (26) aswell as of (17) lead 
to a number of phenomenological implications apart from 
neutrino masses, like new exotic decays or just new important
 contributions to various processes. Most of these have been
 analyzed in the literature\cite{DRG} and will not be considered here.
 In the present article we addressed the question of whether
 it is posssible in GUTs to obtain R-parity violation with 
a large $\Delta L/\Delta B$ hierarchy of strengths so that the proton 
stability is ensured while interesting $L$-non-conserving 
 processes exist at appreciable rates. We considered 
variants of the SU(5) GUT with a built-in Peccei-Quinn 
symmetry suitable for suppressing D=5 $B$-violating operators.
 It turns out that a spontaneously broken Peccei-Quinn
 symmetry in conjunction with an appropriate field 
content can result in an effective R-parity breaking 
characterized by a large hierarchy.

\end{document}